# Exchange Bias following Kinetic Arrest

Praveen Chaddah
Pahle India Foundation, Sarvodaya Enclave, New Delhi 110017, India.
Email: chaddah.praveen@gmail.com

Exchange bias is often observed when anti-ferromagnetic and ferromagnetic phases coexist. The coexistence of two competing magnetic phases can persist to the lowest temperatures if the disorder-broadened 1$^{st}$ order transition separating them is interrupted, as is proposed in the 'kinetic arrest' phenomenon. The fractions of coexisting phases can, in this phenomenon, be tuned by following different cooling protocols. We discuss predicted behaviours of exchange bias resulting from the kinetic arrest phenomenon. Specifically, for appropriate values of cycling field $H_{max}$ and measuring temperature $T_0$ there will be no exchange bias under cooling in zero field, while it will manifest with increasing cooling field, and then saturate.

1. Introduction

Exchange Bias (EB) is a much reported phenomenon [1] in which an isothermal M-H loop is not symmetric about the origin. In most such reports [2,3] it is symmetric (with inversion symmetry) about a point ($M_0$, $H_{EB}$) where $H_{EB}$ is termed as the exchange-bias field . In some recent reports of the EB phenomenon, the isothermal M-H loop does not even display inversion symmetry [4].

There is a belief that exchange bias is observed when anti-ferromagnetic (AFM) and ferromagnetic (FM) phases coexist [1]. Empirically, EB has been reported with ferromagnetic clusters in an anti-ferromagnetic matrix, with ferromagnetic clusters in an anti-ferromagnetic matrix, or even with a ferromagnetic film on an anti-ferromagnetic layer. The EB phenomenon is not yet fully understood, though there is a suggestion [4] that EB is most pronounced with ferromagnetic clusters of small size in an anti-ferromagnetic matrix.

Two competing magnetic phases will coexist at the transition temperature of a first order transition. This coexistence persists over a finite range of temperatures in disorder-broadened transitions. The coexistence of two competing magnetic phases can persist to the lowest temperatures if this disorder-broadened transition is interrupted, as is proposed in the 'kinetic arrest phenomenon'. Since the transition between two competing magnetic phases can be caused both by varying temperature (T) or by varying magnetic field (H), the kinetic arrest phenomenon can be observed by varying either H or T in appropriate regions of (H,T) space. The occurrence of kinetic arrest on cooling in constant H depends on the value of the cooling field; this also dictates the fractions of the two coexisting phases.

Coexistence of two competing magnetic phases is a consequence of kinetic arrest phenomenon, and this coexistence is believed to be a necessary (but not sufficient?) condition for exchange bias. Does the occurrence of phase coexistence following kinetic arrest necessarily result in the observation of exchange bias? In a recent paper Cakir et al [2] found that kinetic arrest and exchange-bias effects occur concurrently in a Ni–Mn–Ga Heusler alloy. Other Ni–Mn-based Heusler alloys also show both exchange bias and the kinetic arrest phenomenon, both in bulk samples [5] and in melt-spun ribbon samples [6]. A large exchange bias was reported [7] in the Heusler compound $Mn_2PtGa$, which was later [8] also found to exhibit the kinetic arrest phenomenon in the appropriate (H,T) range. Is the simultaneous observation of exchange bias and kinetic arrest fortuitous, or is exchange bias a necessary consequence of phase coexistence following the kinetic arrest phenomenon? All the cases mentioned above show a first-order FM (or ferrimagnetic [7]) to AFM transition as T is lowered, and we shall restrict ourselves in this paper to this low-magnetization ground state.

Coexistence of two competing magnetic phases even as the magnetic field is reduced and cycled through H=0, is an essential condition for exchange bias. In the kinetic arrest phenomenon for magnetic first order transitions, the fractions of the two phases coexisting at H=0 can be tuned by following different paths in (H, T) space [9]. Since the fractions of coexisting phases can be tuned in a predictable manner, we should be able to provide measurement protocols to check whether or not the observation of exchange bias is a consequence of the kinetic arrest phenomenon. This could also provide tests on theoretical predictions of how EB varies with cluster size. Finally, since a large number of families of materials show the kinetic arrest phenomenon, we should be able to test if they also show exchange bias under such measurement protocols. This would distinguish between the role of coexisting fractions, and of microscopic magnetic interactions. These are the motivations for this paper.

2. Phase coexistence in kinetic arrest under different protocols

We consider the case where the transition is from a higher temperature ferromagnetic (or even ferrimagnetic) to a lower temperature AFM phase. The schematic in figure 1 shows the broadened H**-T** band corresponding to the limit for superheating (or the superheating spinodal), the broadened H*-T* band corresponding to the limit for supercooling (or the supercooling spinodal), and the kinetic arrest $H_K$-$T_K$ band. This schematic makes the simplifying assumption that the slopes of each of these three bands are independent of H, but the conclusions we shall reach do not invoke this simplification.

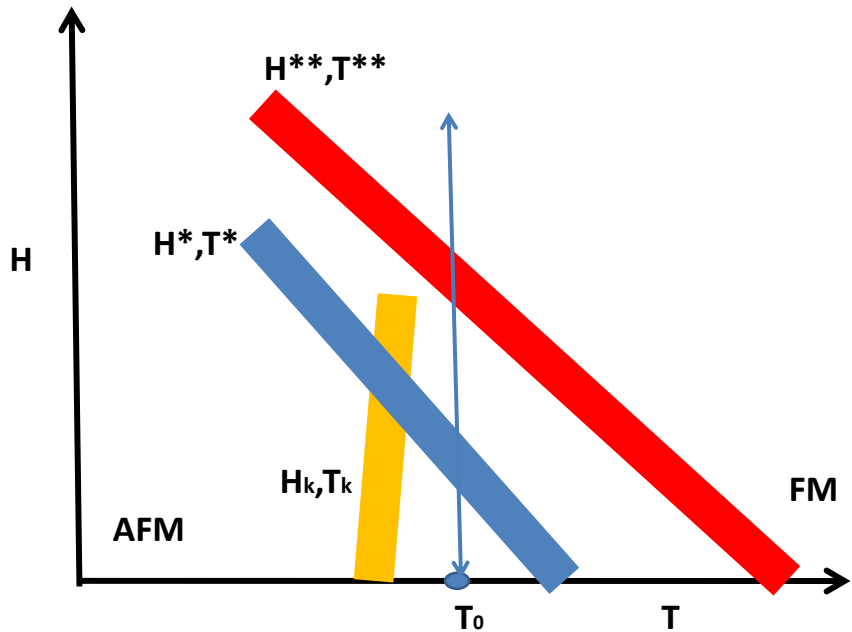

**Figure 1.** Schematic corresponding to $T_O$ lying above the $H_K$-$T_K$ band. Irrespective of the value of $H_{max}$, the material will be in a homogeneous AFM phase at H=0.

2.1 We shall consider different cases depending on the value of $T_0$, the temperature at which isothermal M-H measurements are made. If $T_O$ lies above the $H_K$-$T_K$ band as shown in the schematic figure 1, then any arrested fraction is de-arrested (and there is no kinetic arrest) when H is reduced to zero. Consequently the material is in the homogeneous AFM phase at H=0 and there is no exchange bias for such $T_0$. This is irrespective of whether the material is cooled in zero field, or in some finite value of H.

2.2 We now consider the case where $T_0$ lies within the $H_K$-$T_K$ band at H = 0, as shown in the schematic figure 2. In this case the material will be in AFM phase if it is cooled in H = 0, or if it is cooled in any field lower than $H_2$ and H is then reduced to zero. In these two cases, there will be phase coexistence only if H is isothermally raised to a value $H_{max}$ that lies in the H**-T** band. The fraction of FM phase persisting at H = 0 will rise with increasing $H_{max}$, and will saturate when $H_{max}$ rises above the H**-T** band. This fraction depends on the value of $T_0$. This dependence of FM fraction on $H_{max}$ is depicted schematically in figure 3a. The FM fraction increases monotonically with increasing $H_{max}$.

For a cooling field $H_{cool}$ that lies between $H_2$ and $H_1$, there is an arrested FM fraction in the field-cooled state as H is reduced to zero. This arrested FM fraction rises monotonically from zero to the saturated value (dependent on $T_0$) as $H_{cool}$ rises from $H_2$ to $H_1$. This fraction is independent of $H_{max}$ as long as $H_{max}$ lies below the $H^{**}$-$T^{**}$ band, and depends only on $H_{cool}$. This dependence of FM fraction is depicted schematically in figure 3b.

We consider further the situation where $T_0$ lies in this range, and $H_{max}$ is kept fixed at a value larger than $H_1$ but lies below the $H^{**}$-$T^{**}$ band. In the ZFC case the material is in the homogeneous AFM phase and there is no exchange bias. As the cooling field is raised the same situation persists till $H_{cool}=H_1$, above which the as-cooled material has some arrested FM fraction. A part of this arrested fraction will get de-arrested when H is reduced to zero. As $H_{cool}$ rises to $H_2$ both the arrested FM fraction, and the FM fraction surviving at H=0, rise. Above $H_{cool}=H_2$ this saturates. Consequently, we shall observe $H_{EB}=0$ in the ZFC case, and remaining zero till $H_{cool}=H_1$, rising as $H_{cool}$ rises, and finally saturating above $H_{cool}=H_2$. **We note that such an $H_{EB}$ rising monotonically from zero to a saturated value with increasing $H_{cool}$ has been reported in MnPtGa and MnFeGa materials (See figure 3 and figure S-14 of ref. [4]).**

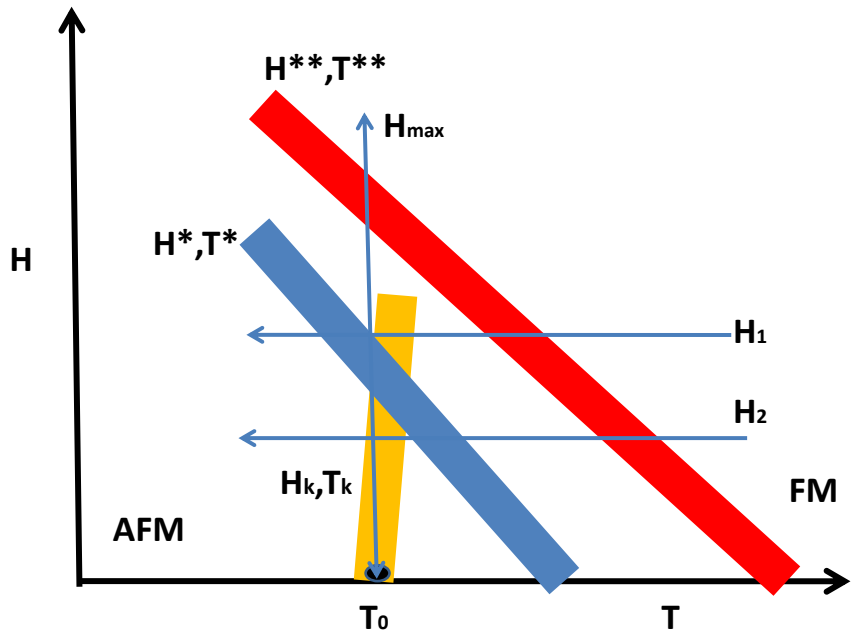

**Figure 2.** Schematic corresponding to the most interesting case of $T_0$ lying within the $H_K$-$T_K$ band at H=0. See text for details.

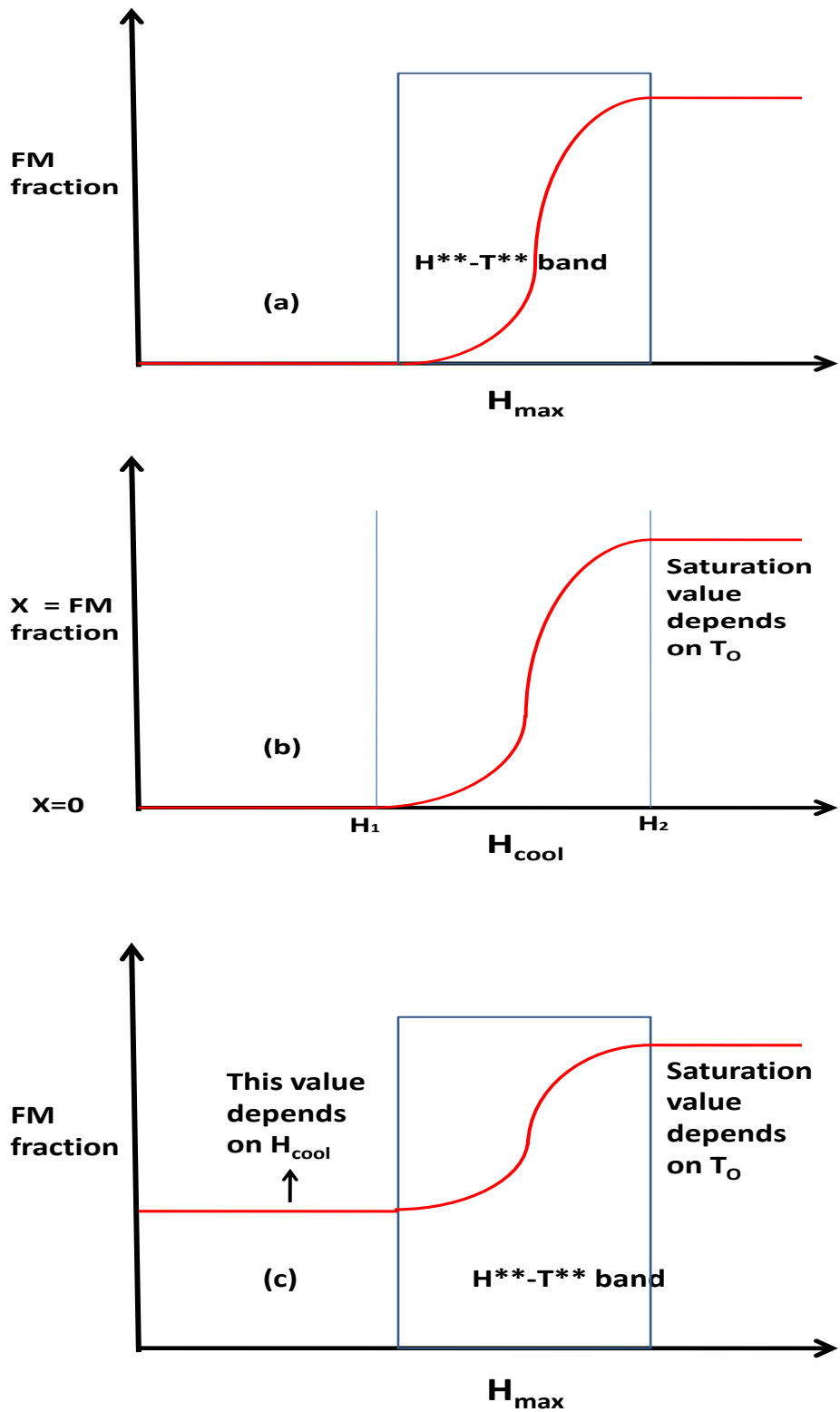

**Figure 3.** Schematic showing FM phase fraction for different histories. See text for details.

If, however, H is isothermally raised to a value $H_{max}$ that lies in the $H^{**}$-$T^{**}$ band, then the fraction of FM phase rises further with increasing $H_{max}$. It will saturate when $H_{max}$ rises above the $H^{**}$-$T^{**}$ band. This saturation value is independent of $H_{cool}$ and depends only on the value of $T_0$. This is depicted schematically in figure 3c, where the lowest value of $H_{max}$ used is, obviously, $H_{cool}$.

We should note here that this region of $H_{max}$ lying in the $H^{**}$-$T^{**}$ band, and this range of $T_0$, lying within the $H_K$-$T_K$ band at H = 0, led to the first report and conjecture of kinetic arrest [11]. It is in this circumstance that one observes the visually striking situation of the virgin M-H curve lying outside (below) the envelope M-H hysteresis loop.

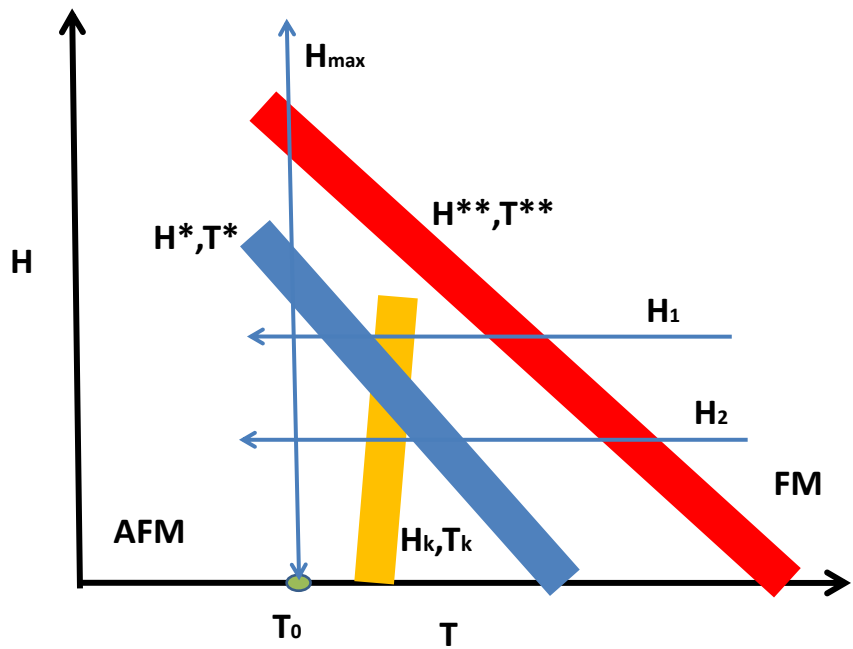

**Figure 4.** Schematic corresponding to $T_O$ lying below the $H_K$-$T_K$ band. See text for a detailed discussion.

2.3 We finally consider the case when $T_0$ lies below $T_K$-$H_k$ band, as shown in the schematic figure 4. We consider that the material is cooled in H = 0, or in some field $H_{cool}$ lower than $H_2$. The sample is now in the equilibrium AFM phase. If H is raised to some $H_{max}$ below the $H^{**}$-$T^{**}$ band, then the sample has remained in the homogeneous AFM phase at $H_{max}$ and remains in this homogeneous phase as H is cycled through zero. There will be no exchange bias because there is no fraction of FM phase, and no phase coexistence, at H=0.

If H is isothermally raised to a value $H_{max}$ that lies in the $H^{**}$-$T^{**}$ band, then there is some conversion to the FM phase and the fraction of FM phase rises with increasing $H_{max}$. The fraction of FM phase reached at $H_{max}$ will remain frozen as H is reduced because T is always below the $H_K$-$T_K$ band. There can be no back-conversion to the equilibrium AFM phase. Exchange bias will be observed until $H_{max}$ rises above H**-T** band. In that situation the sample has converted to a fully ferromagnetic phase at $H_{max}$ and remains arrested in this homogeneous phase as H is cycled through zero. There will now be no exchange bias because there is no phase coexistence at H=0. So, exchange bias is now observed only in a narrow region of $H_{max}$.

For a cooling field $H_{cool}$ that lies between $H_2$ and $H_1$ there is an arrested FM fraction in the field-cooled state. This arrested FM fraction rises from zero to one as $H_{cool}$ rises from $H_2$ to $H_1$. This fraction remains fixed with varying H as long as $H_{max}$ lies below the H**-T** band, and depends only on $H_{cool}$ as depicted in figure 3b. Exchange bias is now observed, and the dependence of FM fraction on $H_{cool}$ is as was depicted schematically in figure 3b. Exchange bias will be not be observed if $H_{max}$ rises above H**-T** band.

## 3    Discussion

We have discussed the behavior of coexisting phases when an FM to AFM transition undergoes kinetic arrest. The resulting phase coexistence provides a necessary condition for the observation of exchange bias. We find, for appropriate values of cycling field $H_{max}$ and measuring temperature $T_0$, a dependence on cooling field that resembles reported observations. We provide detailed verifiable prediction on the qualitative dependence of phase coexistence on cooling histories.

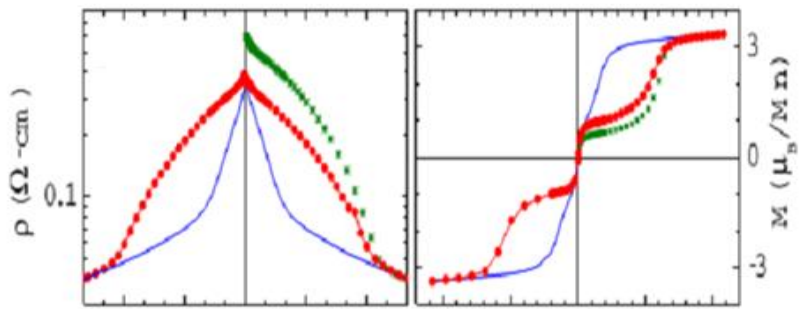

**Figure 5.** Virgin curve is lying outside the hysteresis curve in both resistivity (left) and magnetization (right) measurements under isothermal variation of H. This corresponds to $T_0$ lying in the $H_k$-$T_k$ band. The onset of partial back conversion on reducing H is more apparent in the resistivity measurements.

The most interesting behavior is observed at the same $T_0$ at which one observes the virgin curve lying outside the envelope hysteresis curve in isothermal measurements of magnetization, or of resistance, with cycling H. This happens either when $T_0$ lies below $T_K$-$H_k$ band, or when $T_0$ lies within the $T_K$-$H_k$ band. At the lower $T_0$ the transformed FM phase obtained at $H_{max}$ does not convert back to AFM phase on lowering H. When $T_0$ lies within the $T_K$-$H_k$ band then the transformed FM phase at $H_{max}$ is partially converted back to the starting AFM phase. Since this back conversion happens near H=0, where the FM phase magnetization is small in magnitude, the back conversion is not obvious in M-H measurements. But resistivity in the FM phase does not rise with lowering H, and the back conversion is visually obvious in the return leg of resistivity vs. H, as shown in figure 5 in the representative data from Rawat et al [12]. However, the interesting dependence of $H_{EB}$ on $H_{cool}$ is observed at $H_{max}$ below the H**-T** band, whereas the virgin curve lying outside the hysteresis curve depicted in figure 5 is observed when $H_{max}$ lies within, or above, this band. To test our prediction one has to work at a temperature where the behavior of figure 5 is observed, and then choose a lower $H_{max}$ so that the ZFC M-H does not show this behavior! However, even for this $H_{max}$, the virgin curve of ZFC M-H will lie below the hysteresis curve completed after cooling in a field that lies between $H_1$ and $H_2$, or is higher.